\documentclass[journal,11pt,onecolumn,draftclsnofoot]{IEEEtran}
\usepackage[top=1in, bottom=1in, left=1.25in, right=1.25in]{geometry}
\usepackage{cite}
\usepackage{hyperref}

\usepackage{xspace}
\DeclareRobustCommand\onedot{\futurelet\@let@token\@onedot}
\def\@onedot{\ifx\@let@token.\else.\null\fi\xspace}

\usepackage{placeins}
\usepackage[pdftex]{graphicx}
\ifCLASSINFOpdf
\else
\fi
\usepackage{epsfig}
\usepackage{color}
\usepackage{subcaption}
\usepackage{graphicx}
\usepackage{amsmath}

\begin{document}
\captionsetup[figure]{name={Figure}}

%
\title{Physics-guided Terahertz Computational Imaging}
%
%
%

        
\author{Weng-Tai Su, Yi-Chun Hung, Po-Jen Yu,
    Chia-Wen Lin,~\IEEEmembership{Fellow,~IEEE,}
    Shang-Hua Yang,~\IEEEmembership{Member,~IEEE}}
\maketitle



%
\IEEEpeerreviewmaketitle

Visualizing information inside objects is an ever-lasting need to bridge the world from physics, chemistry, biology to computation. Among all tomographic techniques, terahertz (THz) computational imaging has demonstrated its unique sensing features to digitalize multi-dimensional object information in a non-destructive, non-ionizing, and non-invasive way. Applying modern signal processing and physics-guided modalities, THz computational imaging systems are now launched in various application fields in industrial inspection, security screening, chemical inspection and non-destructive evaluation. In this article, we overview recent advances in THz computational imaging modalities in the aspects of system configuration, wave propagation and interaction models, physics-guided algorithm for digitalizing interior information of imaged objects. Several image restoration and reconstruction issues based on multi-dimensional THz signals are further discussed, which provides a crosslink between material digitalization, functional property extraction, and multi-dimensional imager utilization from a signal processing perspective.

\section{Introduction}
\label{sec:intro}
In recent years, due to the explosive development of information technology, imaging technology has become more and more extensive such as a wide range of applications in consumer electronics, scientific imaging, human-computer interaction, medical imaging, microscopy, and remote sensing. Emphasis is on applied image processing and solving inverse problems using classic algorithms, formal optimization, and modern artificial intelligence techniques. To discover information from invisible to visible, imaging modalities based on different electromagnetic spectrum regimes become an emerging field for establishing digital twins in a virtual world. X-ray imaging is one of the most representative methods to precisely visualize the interior of objects. Despite its capability, high-energy X-ray photons would cause both destructive and ionizing impacts to varieties of object types preventing further investigations with other object characterization modalities. Functional imaging techniques such as Raman spectroscopy \cite{long1977raman}, hyperspectral imaging \cite{lu2014medical}, and temporal imaging \cite{kolner1994space} were further developed for extracting physical nature of an object. Recently, terahertz (THz) imaging technology has attracted significant attention due to its non-invasive, non-destructive, non-ionizing, material-sensitive, and ultrafast dynamics natures for object visualization and exploration. Through THz-matter-interaction, multi-functional tomographic information of the objects can be retrieved even at a remote distance. 

THz spectrum, in between microwave and infrared, has often been described as the last frontier of electromagnetic wave. As THz wave can partially penetrate through varieties of optically opaque objects, it carries hidden material information along the traveling path, making this approach a desirable way to visualize the interior of objects without damaging the exterior \cite{jansen2010terahertz}. Moreover, the rotational, vibrational, torsional frequencies of a great variety of molecules fall in the THz regime. The "self-born spectral barcodes" enrich the retrieved object information with geometrical information and the ingredient information onsite. This contact-free, label-free technique has been proven for studying transport mechanisms of emerging materials with sub-picosecond time resolution \cite{spies2020terahertz}. Thanks to the great advances of recent high-speed electronics, optoelectronics, and thermal electronics technologies, the heart of the THz imaging systems –- THz sources/detectors –-  becomes much more compact, energy-efficient, easy-to-use, and ready for integrating with commercially available on-chip systems. 

In the sub-THz regime, high electron mobility transistor (HMET) and heterojunction field-effect transistor (HFET) have been widely used as the critical component of monolithic microwave integrated circuits (MMIC). Those mm-size, high-speed, efficient THz electronics can offer bright THz radiation reaching above 1 THz. THz photoconductive antenna (PCA) now dominates the market because it can generate/detect coherent electromagnetic signals with an ultrabroad covering range from tens of GHz to tens of THz. Notably, these compact THz optoelectronics can be highly integrated with the matured fiber communication, laser, and semiconductor manufacturing industry, making them highly favorable for modular imaging solutions. Considering real-world application schemes, the unmet data acquisition speed drives the need for miniaturizing a massive amount of THz detectors and device integration in a planar platform. Up to date, Commercially available THz cameras are mainly based on microbolometer array and field-effect transistor (FET) array technologies. These energy-sensitive THz microbolometer cameras consist of 10k--1M sensing elements with a sensor size of tens of micrometer operating at higher than 10 Hz, which is now well-suitable for real-time THz computational imaging applications. Phase-sensitive or spectral-resolved THz imagers are still gradually reaching a level of maturity for scientific instrument use. Therefore, the current approaches to perform hyperspectral THz imaging, time-resolved THz imaging, or THz computed tomography primarily utilize single THz detector matching with different computed imaging modalities. Then here comes a question: \textit{How can we bridge the THz physical properties and the adopted computational imaging methods to extract invisible objection information?}

To answer this question, this work presents a tutorial on state-of-the-art THz computational imaging techniques that combine physical-guided models to achieve high-quality objection information restoration and reconstruction. First, we will focus on the widely used THz computational imaging system configurations that generate the THz dataset in multi-dimensional domains (e.g., time, frequency). By adequately incorporating wave phenomenon into computational imaging methods, we highlight the additional functional imaging and super-resolution features that physical-guided models can offer in recent years. In order to gradually introduce from physics-guided to physics-guided data-driven methods, we introduce on how we benefit from data-driven methods to achieve restoration and reconstruction of THz images measured, especially through the assistance of frequency-domain information which can be obtained from a THz-TDS system. The physics-guided data-driven method captures such complementary spectral characteristics of materials to restore corrupted THz images effectively. Finally, we demonstrate the effectiveness of physics-guided data-driven methods for THz computational imaging through experiments on real-world data.

\section{Conventional THz Computational Imaging}
\label{sec:Conventional}

In the past few decades, most THz images have been retrieved by raster-scanning an object under test through diffraction-limited spots. Depending on the selection of THz detector and how THz radiated waves interact with the illuminated regime, raster-scanned THz imaging systems can reveal plenty of information inside objects, including sub-surface structures of culture heritages \cite{jackson2011survey}, contactless conductivity mapping \cite{lloyd2012review}, defect locations in industrial products, and tumor profile from freshly excised breast tissue \cite{ashworth2009terahertz}. Extended to the raster-scanned configuration, structure-light-based and holographic-based THz imaging systems provide prominent approaches to relief mechanical movement constraints and transfer the data acquisition burden to the computation side. This section will focus on the major THz imaging system configurations to collect object information (Figure 1). 

\begin{figure*}[!hbt]
\centering
\vspace{-0.25in}
\includegraphics[width=1.0\textwidth]{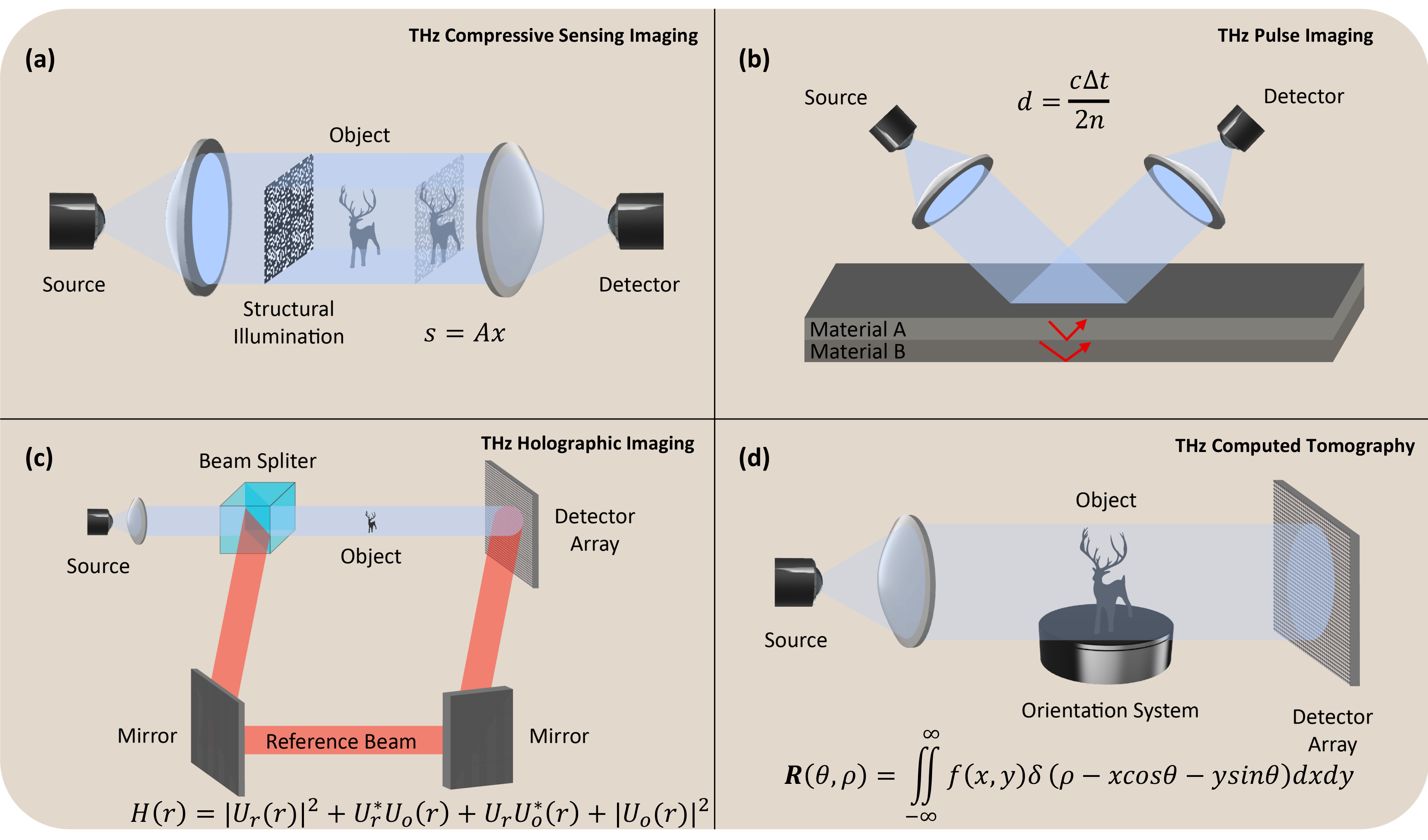}
\vspace{-0.4in}
\caption{\small Illustration of THz computational imaging system configurations: (a) THz compressive sensing imaging, (b) THz pulse imaging, (c) THz holographic imaging, and (d) THz computed tomography.} 
\label{fig:system}	
 \vspace{-0.2in}
\end{figure*}

\vspace{-0.2in}
\subsection{THz Compressive Sensing Imaging}

Instead of raster-scanning objects through either moving them or steering the light path, compressive sensing (CS) imaging is a computational method to reconstruct object information with low-rank detection. Depending on the information sparsity of the objection and CS modalities, the number of measurements can be greatly reduced compared to the pixel number. This brings three major advantages to the THz imaging system configuration –- high-speed data acquisition, the use of a single THz detector, no mechanical moving parts –- at the expense of increased computational effort. The basic operation principle of THz CS imaging is to illuminate a series of patterned THz beams on the object. The multiplication of pixelated masks and vectorized object geometry then record the total transmitted THz signal by a single-pixel (bucket) THz detector as described in the following equation.
\begin{align} \label{eq:compressed_sensing}
    \mathbf{s} = \mathbf{A}\mathbf{x},
\end{align}
where $\mathbf{s}$, $\mathbf{A}$ and $\mathbf{x}$ are the recorded THz signal, the pixelated mask and the vectorized image to be reconstructed, respectively. Each row of the sensing matrix represents an illumination pattern. To solve (\ref{eq:compressed_sensing}), the following optimization problem can be used.
\begin{align}
    \min_\mathbf{x} ||\mathbf{s}-\mathbf{A}\mathbf{x}||_2 + \lambda ||\mathbf{x}||_1,
\end{align}
where $\lambda$ is a hyperparameter and the regularizer $||\mathbf{x}||_1$ is used to promote the sparsity of the solution $\mathbf{x}$. 
A great variety of high-speed, high-modulation-depth THz spatial light modulators has been invented in the past ten years. With the combination of photoconductive modulation and digital light processing techniques, a real-time THz video (frame rate: 6 fps; pixel number: 32 x 32) has been demonstrated \cite{stantchev2020real}.

\vspace{-0.2in}
\subsection{THz Pulsed Imaging}

Pulsed imaging, or time-of-flight imaging, records the traveling THz pulsed signal starting from the THz source, tested sample to the THz detector. A typical system configuration is based on a THz time-domain spectroscopy (THz-TDS) system  performed in the reflection mode. The time-delayed signals and the echoes reflect inner object structures.
\begin{align}
    d = \frac{c\Delta t}{2n}, \nonumber
\end{align}
where $d$ is the distance from sample surface to the sample internal interface; and $c$, $n$ and $\Delta t$ are the speed of light in vacuum, the sample refractive index, and the time delay of the recorded signal.
While conventional THz-TDS systems suffered from long data acquisition time due to mechanical components, modern techniques such as asynchronous optical sampling THz-TDS systems and electronically controlled optical sampling systems significantly speed up data acquisition time with orders of magnitude -- from hours to milliseconds -- bringing THz pulsed imaging modality to become a unique but powerful tool to discover multi-functional material properties in wide ranges of spatial, temporal, spectral and material domains, simultaneously. To this extent, THz pulsed imaging has been allocated for many applications including drug detection and compressed sensing imaging.

\vspace{-0.2in}
\subsection{THz Holographic Imaging}

Inspired by optical holography and synthetic aperture radar technologies, the THz holographic imaging system not just records the transmitted THz beam profile but the phase changes through the imaged objects \cite{heimbeck2020terahertz}. This recorded THz amplitude/phase mapping represents the thickness, topology, features, and dielectric response of the object, which can further reconstruct the whole object geometry beyond diffraction-limited spatial resolution. A typical THz holographic imaging system is based on a Mach-Zehnder THz interferometer setting --- the superposition of the coherent scattered THz beam and the reference THz beam --- as described in equation \eqref{eq:THz_holo}.
\begin{align} \label{eq:THz_holo}
    \mathbf{H}(r) = |\mathbf{U_r}(r)|^2 + \mathbf{U_r}^* \mathbf{U_o}(r) + \mathbf{U_r} \mathbf{U_o^*}(r) +|\mathbf{U_o}(r)|^2,
\end{align}
where $\mathbf{H}(r)$, $\mathbf{U_r}(r)$, and $ \mathbf{U_o}(r)$ are the THz interferogram, the electric field of the reference beam, and the electric field of coherent scattered THz beam from the object, respectively.
Recent trends in time-domain THz holography, one-shot THz holography, and video-rate THz digital holography have arisen extensive attention in this field. These modalities extend the system capability to functional 3D tomography at applicable frame rates. With high-fidelity amplitude and phase reconstruction, the depth resolution can reach 2.2 $\mu m$ ($\lambda/284 $ at 0.48 THz), comparable with the conventional optical imaging system.

\vspace{-0.2in}
\subsection{THz Computed Tomography}

As the analogy of X-ray computed tomography (CT), a typical THz CT system is composed of a THz source-detector pair and an object positioning stage that can rotate and move in the local coordinate system. While the THz beam transmits through the objects’ local regime, the loss function of the THz signal, $f(x,y)$, then be recorded by the THz detector in use. The data content of recorded signals could be field strength, spectral amplitude/phase, or polarization changes. By sampling from different angular and radial positions, one can define the function $R(\theta, \rho)$ , the Radon transform, to relate the object information with the recorded signals.
\begin{align}
    \mathbf{R}(\theta, \rho) = \int_{-\infty}^{\infty}\int_{-\infty}^{\infty}f(x,y)\delta(\rho-x\cos\theta-y\sin\theta)dxdy, \nonumber
\end{align}
where $\theta$ and $\rho$ are the angular and radial coordinates of the projection line $(\theta, \rho)$, respectively; and $\delta(\cdot)$ is the Dirac impulse. The Inverse Radon Transform modal through filtered back-projection \cite{kak2001algorithms} or iterative methods (i.e., simultaneous algebraic reconstruction technique, expectation-maximization) defines the way to reconstruct each object voxel and its physical contents from the projected dataset. In a practical scenario, the static/dynamic measurement misalignments and the error from data discretization significantly impact the reconstructed data, which is a sparse research field waiting to be explored.

\begin{figure*}[!hbt]
\centering
\includegraphics[width=1.0\textwidth]{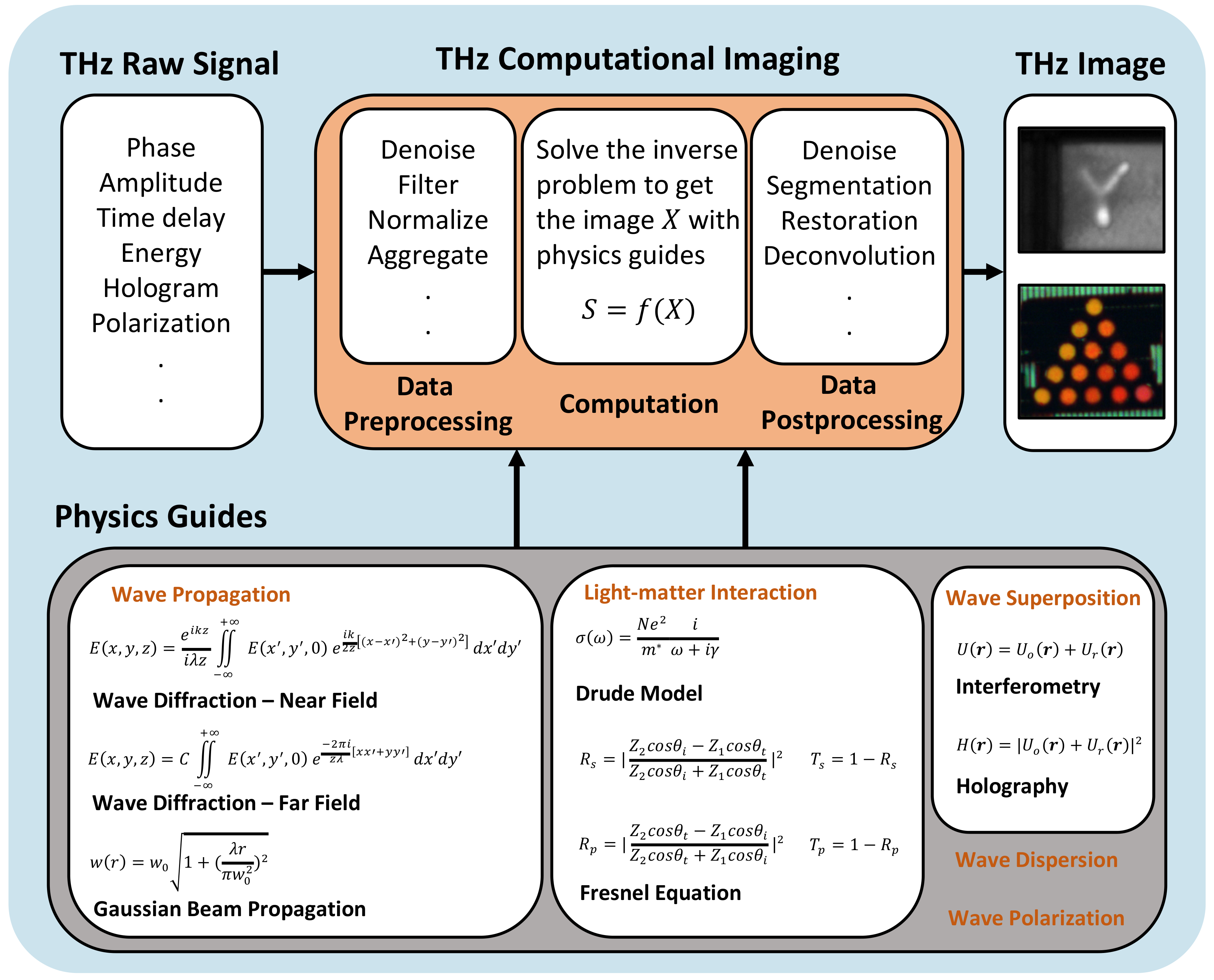}
\vspace{-0.4in}
\caption{\small Flowchart of physics-guided THz computational imaging.} 
\label{fig:THz_compu_flow}	
\vspace{-0.2in}
\end{figure*}

\begin{figure*}[!hbt]
\centering
\includegraphics[width=0.9\textwidth]{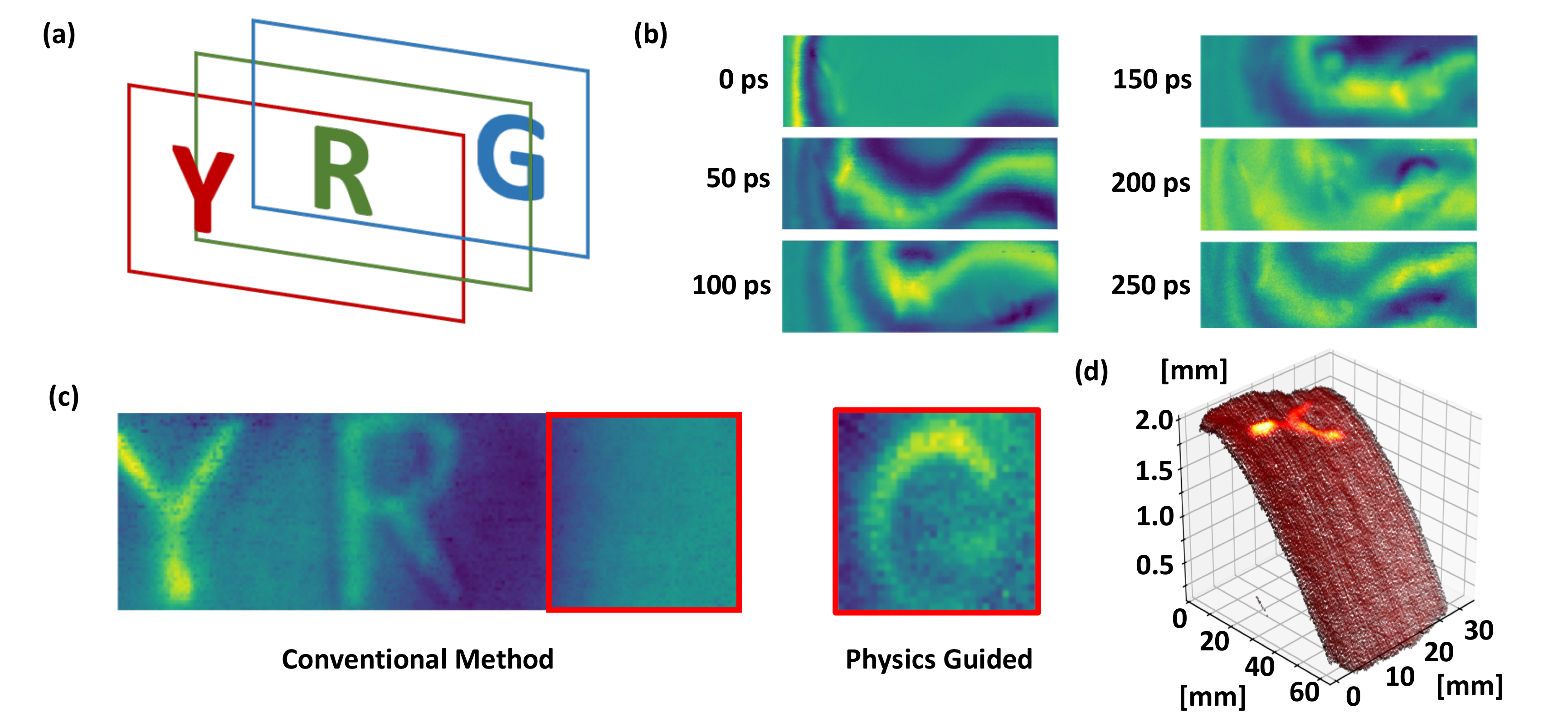}
\vspace{-0.25in}
\caption{\small An example of THz computational imaging based on a time-of-flight system. (a) The illustration of the measured object composed of three paper sheets with letters ``Y'', ``R'' and ``G'', respectively; (b) The THz electric field mapping at different time stamps. (c) The images retrieved by a conventional method and a physics-guided method, respectively. (d) The reconstructed 3D image of the first paper sheet by using the time-of-flight information.} 
\label{fig:THz_compu_flow2}	
\vspace{-0.3in}
\end{figure*}

\section{Physics-guided THz Imaging}
\label{sec:guided_conv}
Many THz imaging modalities have been developed based on the fundamental properties of THz wave – amplitude, phase, spectrum, polarization, wavefront, chirality, coherence, wave propagation, and light-matter interactions \cite{guerboukha2018toward}. Physics-guided methods are commonly used in the sequential stages of data pre-processing, image computation, and image post-processing to translate the object information from the raw data (Figure 2). Although most THz computational imaging studies are based on specific data types, it should be noted that the fusion of multiple data types, such as spatio-temporal-spectral data, can further enhance image quality. With the utilization of prior knowledge (\textit{e.g.}, beam profile, spectral fingerprints, and ultra-fast light-matter interaction), recent advance in physics-guided THz imaging have shown extraordinary features such as non-invasive carrier dynamics mapping and material inspection on a nanoscopic scale~\cite{aghamiri2019hyperspectral}.

\vspace{-0.2in}
\subsection{Image Formation}
Unlike visible light, most THz imaging systems deal with a spatial resolution close to or shorter than the beam wavelength. As the ray optics can no longer explicitly express the diffraction-limited THz beam profile landing on the object, wave optics play essential roles in THz image reconstruction. The mathematical modeling of THz wave propagation based on Gaussian beam theory profiles the spatially varying beam shapes as the point spread function (PSF) along the traveling axis. With accurate approximate absorption coefficient of the object, beam divergence, and depth of focus, the deconvoluted image's spatial resolution can reach sub-wavelength scale. Phase characteristic of the THz beam is another powerful tool to beat the diffraction limit. The superposition of two coherent THz collimated beams, the planar reference beam and the object beam, forms spatial modulated fringe/speckle patterns that record the optical path differences introduced by the object. Different approaches of scalar diffraction theory (e.g., Fresnel diffraction, Huygens convolution, Rayleigh–Sommerfeld diffraction), which are applicable to in-line and off-axis holography, have been used for reconstructing object 3D information. As the image resolution is mainly determined by the detector aperture, hologram size, and off-axis angle, it is possible to achieve a phase resolution of $\lambda/10000$ – a nano-scale axial resolution using THz beam \cite{heimbeck2020terahertz}. Fresnel equations formulate the light-matter interactions on the material interface and inside the object, which is a well-known approach to precisely reconstruct THz images. With the measurable dielectric response prior to imaged object, it can nicely resolve the thickness variations of multiple stacking layers and the defected regions inside. Along with the combination of the Drude model, which extracts complex conductivity from the material dielectric response, the electrical-property mapping inside the object can be further visualized. By capturing THz light-matter interaction in the near-field regime, research works have been demonstrated object information reconstruction based on Fresnel diffraction in the microscopic world. It is still feasible to put down nanoscale THz imaging by introducing a metallic scattering tip that intensifies the local THz electric field in several orders of magnitude near the tip region. Depending on the light-matter interaction properties of the observed object, single-molecule ultrafast motion can be recorded by THz near-field imaging with sub-molecular resolution \cite{cocker2016tracking}. Moreover, data acquisition speed is an ever-lasting issue for real-time THz imaging. Recent compressive sensing imaging modalities quickly adapt to many THz imaging systems, including THz CT, THz pulsed imaging, THz holography, and more system configurations, which maintain the same object visualization features (e.g., hyperspectral, time-resolved, near-field imaging) with less measurement time \cite{zanotto2020single}. To mitigate the diffraction effect of the reconstructed image caused by the distance between image plane and object plane, the physics guide of inverse Fresnel diffraction can be utilized \cite{shang2019terahertz} with the inverse problem of the THz compressive sensing.

\vspace{-0.2in}
\subsection{Image Pre-processing and Post-processing}
Like in other EM bands, the pre-processing and the post-processing in THz computational imaging can adopt most of the methods developed in signal processing and image processing. The subspace projection on the detected THz spectra achieves the noise removal and, in the meanwhile, preserves the THz spectral features \cite{lin2021hyperion}. The statistics-based average over the THz time-resolved signals can be used to suppress the various types of noises \cite{skorobogatiy2018statistical}. The frequency and wavelet filters are also applied to the THz signal to achieve the noise removal \cite{chen2010frequency}. In the image domain, an isotropic diffusion algorithm and a mixture of Gaussian densities are utilized to achieve the denoising and image segmentation, respectively \cite{shen2008detection}. Based on the rapid development of learning-based approaches in visible light imaging, several deep learning models are also demonstrated in THz computational imaging \cite{park2021machine}. To date, THz imaging processing techniques have less coupled with physics-related working principles. This could be an exciting playground to extend the further capability of THz computational imaging from the signal processing perspective.  

\vspace{-0.2in}
\subsection{Energy-based THz Imaging}
The energy loss encoded in the reflected/transmitted THz signals can distinguish interior object information based on the Fresnel equation; through THz direct detection, energy-based THz imaging is highly favorable for functional 2D imaging or THz CT. The current works focus on feature extraction methods rather than reconstruction models since different feature extraction approaches can directly reflect corresponding material characteristics. For example, the attenuation spectra of THz spectral amplitude signals directly map to the material conductivity locally. While scanning across the whole field of view, the spatial distribution of conductivity in materials can be easily investigated. In \cite{kawase2003non}, Kawase et al. captured multi-spectral THz images for identifying illicit drugs. The selected THz frequencies are based on distinctive spectral fingerprints of each chemical. This way, the spatial distributions and concentration levels of different chemicals can be nicely classified and revealed from THz spectral amplitude signals. The examples above demonstrate that one can directly extract various material information through different THz signal features. However, the energy-based methods suffer a lot from several issues in THz images, such as blurs, noises, low contrast, low image resolution, resulting in difficulties for effective material analysis. To extend the practical use of THz imaging, research groups have been looking for frequency-resolved and time-resolved THz imaging modalities that exploit additional light-matter interaction information to improve image qualities and functional imaging capabilities.

\vspace{-0.2in}
\subsection{Frequency-resolved THz Imaging}
Compared with THz direct detection, THz coherent detection offers not just extremely high dynamic range and low noise equivalent power but the ability to resolve the amplitude/phase spectrum of the THz signals. The amplitude and phase dynamics correspond to the optical delay path pass through the imaged object and its material properties (e.g., complex permittivity), which makes the THz coherent imaging highly favorable for high-precision 3D imaging. Under a swept-source optical coherence tomography (OCT) configuration, the material spectral response and the 3D object geometry can be easily reconstructed based on the optical frequency-domain reflectometry (OFDR) model \cite{nagatsuma2014terahertz}. Typically, THz photomixers and THz quantum cascade lasers (QCLs) are the most promising candidates as the swept sources of THz OCT systems due to their excellent phase-stability, ultrabroad frequency tuning range, fast frequency-sweeping rate, and compact size. To achieve better-reconstructed imaging quality, the modeling of wave propagation, wavefront profile, and dielectric response at each frequency shall be incorporated in the OFDR model. It should be noted that, by synthesizing the broad-frequency-sweeping THz signals, a THz OCT system can form high-quality THz pulses used for time-of-flight THz imaging. 

\vspace{-0.2in}
\subsection{Time-resolved THz Imaging}
Time-resolved THz imaging provides meta-information of imaged objects simultaneously, including time-of-flight, ultrafast phenomenon, charged carrier dynamics and distributions, material spectral fingerprints, and object topology. While the THz pulses pass through the object under evaluation, the reflected time-resolved THz electric field signals from incident inner shells arrive at the THz detector at different timestamps (Figure 3), revealing the geometric information of the object \cite{guillet2014review}. Through FFT operation, the amplitude/phase spectral responses can be further extracted to reveal the objects’ material information. To date, researchers further extend the capability of time-resolved THz spectroscopy (TRTS) systems to visualize the ultrafast dynamics inside photoconductive objects \cite{prabhu1997carrier}. The microscopic electrical properties like conductivity, mobility, and doping mapping can be further derived through the Drude model. While a TRTS system combines with a scattering-type scanning near-field optical microscopy (s-SNOM) system, the time-resolved THz hyperspectral nano-imager can non-invasively resolve nanoscale carrier profiling inside advanced materials \cite{aghamiri2019hyperspectral}. To achieve a better data acquisition rate, Razzari et al. demonstrated a time-domain THz CS imaging system that offers time-resolved, hyperspectral features without the need for mechanical raster-scanning \cite{zanotto2020time}. As TRTS is widely used in pure source identification, Yang et al. developed a blind hyperspectral unmixing method – Hyperspectral Penetrating-type Ellipsoidal Reconstruction (HYPERION) – for THz blind source separation \cite{lin2021hyperion}. This work shows an unprecedented chemical blind separation mapping capability of time-resolved THz hyperspectral imaging without collecting extensive data or sophisticated model training.

\section{Physics-guided Data-driven THz Computational Imaging}
\label{sec:guided}
Many imaging methods have been developed based on the light-matter interaction in the THz frequency range in the past decades. Based on THz absorption imaging modalities, the material complex refractive index mapping can be profiled through Fresnel equation \cite{born2013principles}, extracted by the THz power loss while propagating through the tested object boundary. However, the application scopes of those physics-based methods are severely limited since they normally require a sufficient amount of prior knowledge about a measured object to simplify the complex physical models. To break this limitation, data-driven approaches, especially deep neural networks, have been attracting intensive attention due to their excellent learning capability. 

    Data-driven methods are mainly based on deep learning models \cite{zhang2017beyond}, which do not resort to any explicit transform model but are learnt from representative big data. We can cast THz image analysis as an image-domain learning problem ($\textit{e.g.}$, the THz signals in Figure~\ref{fig:time_freq}). Deep learning has revolutionized the aforementioned physics-based paradigm in image restoration, for which the data-driven methods can be regardless of physical properties while maintaining the advantages of the physics-based methods and achieve state-of-the-art performances. A delicate data-driven model based on physical priors can effectively loosen the requirement of prior knowledge of materials and perform superior to conventional physics-based methods. Nevertheless, THz images retrieved by the conventional physics-guided THz imaging methods do not carry enough restoration information, thereby limiting the efficacy of the data-driven methods. To address the issue, we employ asynchronous optical sampling THz time-domain spectroscopy system (ASOPS THz-TDS) to leverage additional curated pixel-wise spectral information carried in the THz time-resolved signals, such as the prominent amplitude/phase spectral information corresponding to specific light-matter interaction characteristics of THz waves passing through materials. Moreover, data-driven models can excellently fuse the different information of THz signals, such as amplitude/phase spectra and the time-resolved THz signals, to achieve superior image restoration \cite{su2021seeing}. 
    In this section, as an example, we demonstrate the capability of physics-guided data-driven  computational imaging on boosting the performance of an ASOPS THz-TDS CT system.

\vspace{-0.2in}
\subsection{Properties of Time-resolved THz CT Imaging System}
\label{sec:thz-tds}
\begin{figure*}[t]
 \vspace{-0.2in}
 \centering
 \includegraphics[width=0.9\textwidth]{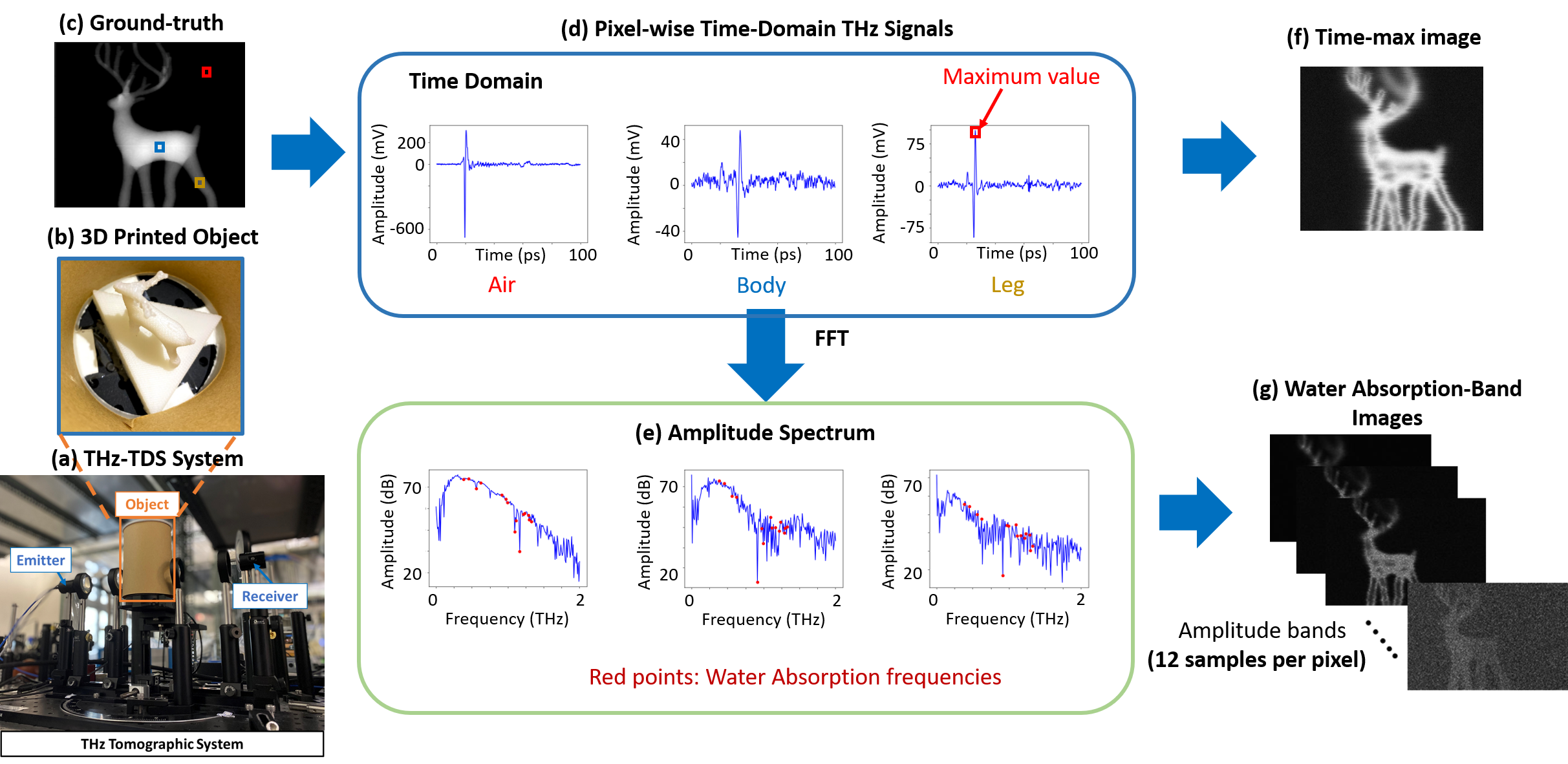}
  \vspace{-0.2in}
 \caption{\small Raw data of THz time-domain data measured in air and the body and leg of a 3D printed deer. (a) The ASOPS THz-TDS CT system; (b) the 3D printed object, (c) the ground-truth of the frontal projected view; (d) the time-domain THz signals of three different pixels (in air and the body and leg of the deer); (e) the amplitude spectra of the three signals; (f) the time-max image (the maximal values of THz time-domain signals in (d); (g) the images retrieved at the water-absorption frequencies.}
 \label{fig:time_freq}	
\vspace{-0.2in}
 \end{figure*}

To measure the THz tomographic images of an object, as illustrated in Figure~\ref{fig:time_freq}(a), we build up an ASOPS THz-TDS CT system prototype. The system is composed of two asynchronous femtosecond lasers with a central wavelength of 1560 nm and a power level of 40mW, a pair of THz photoconductive antenna source and detector, a linear and rotation motorized stage, four plane-convex THz lens with 50 mm focal length, a transimpedance amplifier, and a unit of data acquisition and processing. The repetition rates of the two asynchronous femtosecond lasers are 100 MHz and 100 MHz + 200 Hz, respectively. With the configuration above, our ASOPS THz-TDS system delivers 0.1 ps temporal resolution and the THz frequency bandwidth of 5 THz. Additionally, it provides THz pulse signals with 41.7 dB dynamic range from 0.3 THz to 3 THz and 516 femtoseconds at full width at half maximum (FWHM). 

In our THz-TDS CT system, each object (\textit{e.g.}, the covered 3D printed deer in  Figure~\ref{fig:time_freq}(b)) is placed on the motorized stage in the THz path between the light source and detector of the THz-TDS system. The sample objects are generated by a Printech 3D printer, using the material of high impact polystyrene (HIPS) due to its high penetration of THz waves. Raster scans are performed on the measured object in multiple view-angles, covering a rotational range of 180 degrees (step-size: 6 degrees between two neighboring views),  a horizontal range of 72mm (step-size: 0.25mm), and a variable vertical range corresponding to the object height (step-size: 0.25mm). In this way, we obtain 30 projections of each object, which are then augmented to 60 projections by horizontal flipping. The ground-truths of individual projections are obtained by converting the original 3D printing files into image projections in every view-angle, as shown in Figure~\ref{fig:time_freq}(c).  

During the scanning, the THz-TDS CT system profiles each voxel's THz time-resolved signal with 0.1 ps temporal resolution, whose amplitude corresponds to the strength of THz electric field. Based on the dependency between the amplitude of a time-resolved signal and THz electric field, in conventional THz imaging, the maximum peak of the signal ($\texttt{Time-max}$~\cite{hung2019terahertz}) is extracted as the feature for a voxel. The retrieved $\texttt{Time-max}$ images, as illustrated in Figure~\ref{fig:time_freq}(f), can deliver great SNR and clear object contours. However, the conventional THz imaging based on $\texttt{Time-max}$ features suffers from several drawbacks, such as the undesired contour along the object boundaries, the hollow and holes in the body region, and the blurs in high spatial-frequency regions. To address this limitation, the spectral information of THz temporal signals can be utilized to supplement the conventional method based on $\texttt{Time-max}$ features since the voxel of the material behaviors are encoded in both the phase and amplitude of different spectral bands, according to the Fresnel equation~\cite{dorney2001material}.

Figure~\ref{fig:time_freq}(d) illustrates time-domain THz signals measured in air, the body and leg of a 3D printed deer, respectively. While the THz beam passing through the object, the attenuated THz time-domain signal is encoded with the thickness and material information of the THz-illuminated region. By processing the \texttt{Time-max} images captured at different view-angles, the 3D profile of the printed deer can be further reconstructed. Although this conventional way is well-fitted for visualizing 3D objects, the inherent diffraction behavior and strong water absorption nature of THz wave induce various kinds of noise sources as well as the loss of object information. This leads to the undesired blurring, distorted, speckled phenomenon of functional THz images. Existing works have tackled this issue to restore clear images via estimating point spread functions \cite{popescu2010point, popescu2009phantom}, image enhancement \cite{wong2019computational}, machine learning \cite{ljubenovic2020cnn, wong2019training}, etc. Their performance is, however, still severely constrained by diffraction-limited THz beams.  

 \begin{figure*}
 \centering
  \vspace{-0.25in}
 \includegraphics[width=1.0\textwidth]{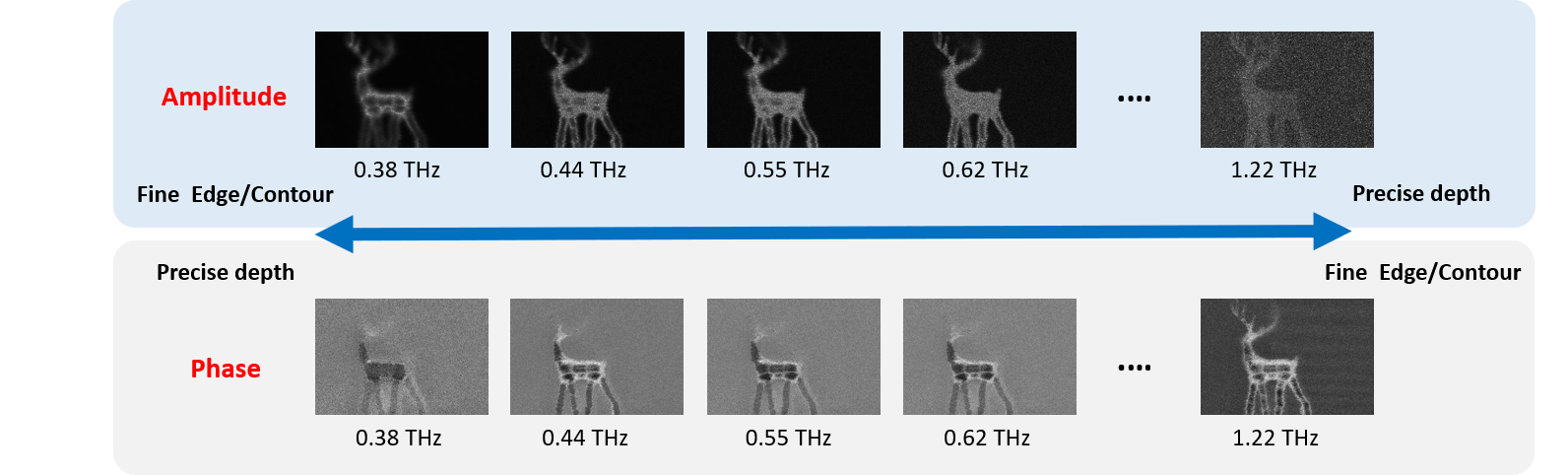}
 \vspace{-0.35in}
 \caption{\small Illustration of THz multi-spectral amplitude and phase images measured from \textbf{Deer} object.} 
 \vspace{-0.2in}
 \label{fig:band}	
 \end{figure*}

\vspace{-0.2in}
\subsection{Physical Guides for THz CT Imaging}
\label{sec:properties_w}

To break the limitations mentioned above, we show that curated physical guides can help extract prominent spatio-spectral information of hidden objects to facilitate the learning of data-driven models so as to boost the performance of reconstructing deep-subwavelength tomographic images of hidden objects. For example, since travelling THz beams can be significantly attenuated at water absorption frequencies, reconstructed THz images based on water absorption lines offer better details on contoured regimes and depth.
As shown in Figure~\ref{fig:time_freq}, each 2D THz image is composed of an array of THz time-resolved signals with a temporal resolution of 10 fs. By taking Fourier transform on the THz time-resolved signals individually, we can extract their pixel-wise multi-spectral feature maps.
Nevertheless, due to the large number of spectral bands with the measured THz data, it is required to sample a subset of the spectral bands to reduce the amount of model parameters. Besides, the THz-TDS system offers relatively better SNR levels in a frequency range of 0.3 THz--1.3 THz. Considering the water absorption in THz regime \cite{van1989terahertz, slocum2013atmospheric} and the superior SNR in the range of 0.3 THz--1.3 THz, we select 12 water absorption frequencies as indicated partly in Figure~\ref{fig:band}.
Prominent multi-spectral information including both amplitude and phase at the selected frequencies is curated and then employed to restore clear 2D images. Specifically, Figure~\ref{fig:band} illustrates multiple 2D THz images of the same object retrieved at the selected frequencies, showing very different contrasts and spatial resolutions as these hyperspectral THz image sets have different physical characteristics through the interactions of different THz bands with an object. 
Specifically, the lower-frequency phase images offer relatively accurate depth information due to their higher SNR level, whereas the higher-frequency phase images offer finer contours and edges because of the shrinking diffraction-limited wavelength sizes (from left to right in Figure~\ref{fig:band}). The phase also contains, however, a great variety of information of light-matter interaction that could cause learning difficulty for the image restoration task. 
%
To address this issue, we utilize selected amplitude bands as complementary information. Although the attenuated amplitude bands cannot reflect comparable depth accuracy levels as phase bands, the amplitude bands still present superior SNR and more faithful contours such as the location information of a measured object. 

In summary, the amplitude complements the shortcomings of phase at the water absorption bands. As a result, fusing the amplitude and phase images from low-frequency to high-frequency brings the following advantages. Since the low-frequency THz signal provides precise depth (the thickness of an object) and fine edge/contour information in the phase and amplitude, respectively, they together better delineate and restore the object. In contrast, the high-frequency feature maps of amplitude and phase respectively provide better edges/contours and precise position information, thereby constituting a better object mask from the complementary features. With these spectral properties of THz images as physical guides, we can extract rich information from a wide spectral range in the frequency domain to simultaneously restore the 2D THz images without any additional computational cost or equipment, which is beneficial for further development of practical THz imaging tasks.

\vspace{-0.2in}
\subsection{CNN-based THz Image Restoration Guided by Water Absorption Profile}
\label{sec:SARNet}

As different EM bands interact with objects differently, THz waves can partially penetrate through various optically opaque materials of objects  and carry hidden object tomographic information along the traveling path. This unique feature
can guide a new approach to visualize the essence of 3D objects, which other
imaging modalities cannot achieve. 
However, the $\texttt{Time-max}$ images in Figure~\ref{fig:time_freq}(f) only preserve energy distribution information which is not enough for learning an effective data-driven model for detailed object information restoration. To this end, we can extract rich information from a wide spectral range in the frequency domain to simultaneously restore the 2D THz images without any additional computational cost or equipment, which is beneficial for further development of THz imaging. The key idea is to fuse spatio-spectral features with different characteristics on a common ground via data fusion approaches to guide the feature fusion for fine restoration. Moreover, we found that directly learning from the full spectral information to restore THz images usually leads to an unsatisfactory performance. The main reason is that the full spectral bands of THz signals involve diverse characteristics of materials, noises, and scattered signals, causing difficulties in the model training of DNNs. Instead of learning from the full spectral information of THz signals, We tackle this problem by extracting prominent multi-spectral information from the complementary amplitude and phase of THz signals in curated water absorption bands.

\begin{figure*}[t]
\centering
\vspace{-0.3in}
\includegraphics[width=0.9\textwidth]{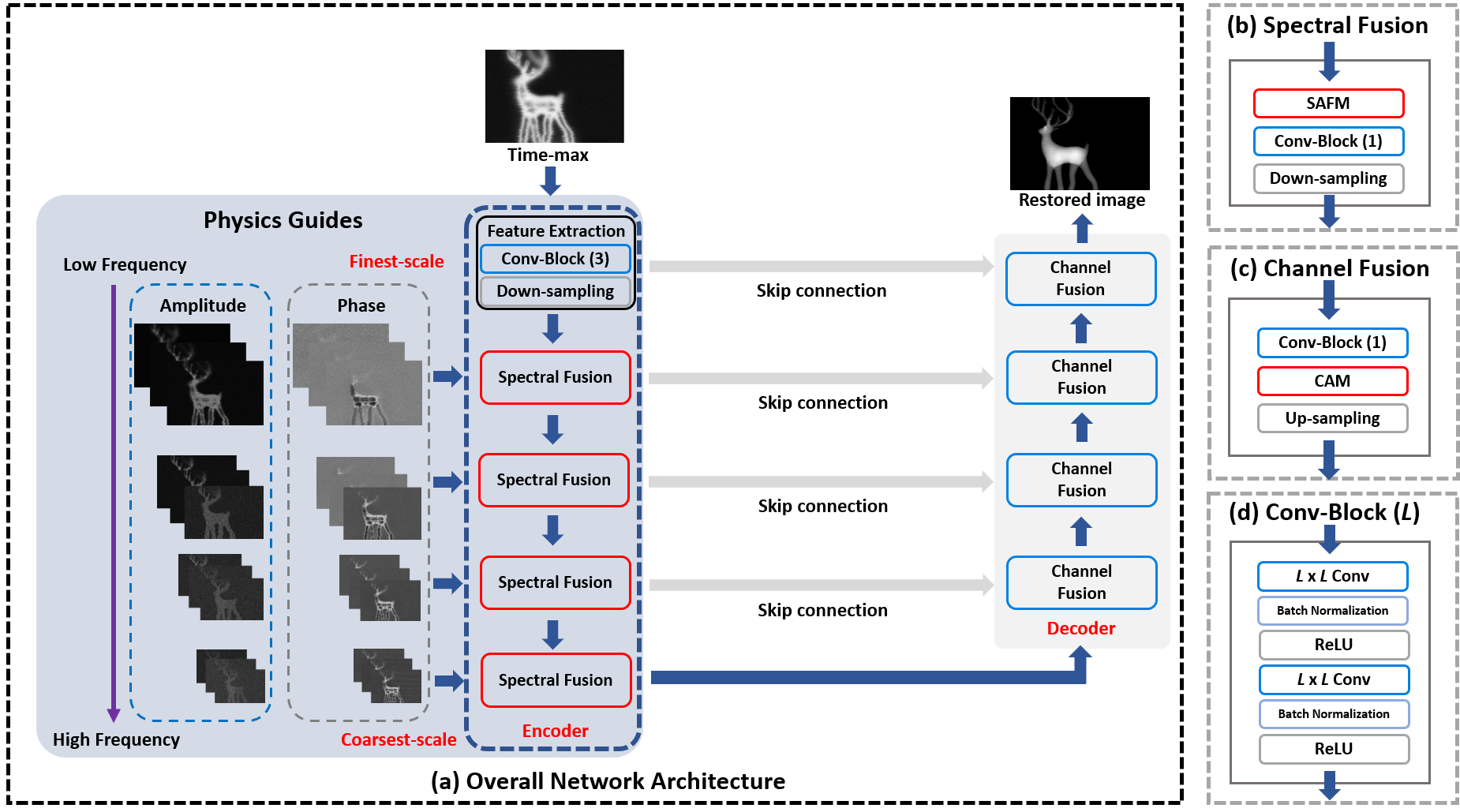}
\vspace{-0.15in}
\caption{\small Network architecture of \texttt{SARNet} consisting of five scale branches, where the finest-scale scale takes the \texttt{Time-max} image as input, and each of the second to fifth takes 6 images of spectral frequencies (3 bands in amplitude and 3 bands in phase) as inputs. The three gray blocks show the detailed structures of (b) Spectral Fusion, (c) Channel Fusion, and (d) Conv-Block.} 
\label{fig:unet}	
\vspace{-0.3in}
\end{figure*}

To accommodate the curated multi-spectral amplitude and phase bands for learning the THz image restoration model in a multi-scale manner, as depicted in Figure~\ref{fig:unet}, we devise a \textbf{S}ubspace-\textbf{A}ttention guided \textbf{Net}work (\texttt{SARNet}) based on the multi-scale U-Net in~\cite{ronneberger2015u}.  Specifically, \texttt{SARNet} is composed of an encoder (spectral-fusion module) with 5 branches of different scales (from the finest to the coarsest) and a decoder (channel-fusion module) with 5 corresponding scale-branches. The spectral fusion block in each scale-branch of the encoder involves a Subspace-Attention-guided Fusion module (SAFM), a convolution block (Conv-block), and a down-sampler, except for the finest-scale branch that does not employ SAFM. To extract and fuse multi-spectral features of both amplitude and phase in a multi-scale manner,  the encoder takes the \texttt{Time-max} image as the input of the finest-scale branch  and  receives to its second to fifth scale-branches 24 images of selected predominant spectral frequencies extracted from the pixel-wise THz signals, where each branch takes 6 images of different spectral bands (3 bands of amplitude and 3 bands of phase) to extract learnable features from these spectral bands. To reduce the number of model parameters, these 24 amplitude and phase images (from low to high frequencies) are downsampled to 4 different resolutions and fed into the second to fifth scale-branches in a fine-to-coarse manner as illustrated in Figure~\ref{fig:unet}. Since the multi-spectral amplitude and phase feature maps are significantly different in nature, the proposed SAFM fuses these features via  firstly learning a common latent subspace shared between the amplitude and phase features to facilitate associating the short/long-range amplitude-phase dependencies. Projected into the shared latent subspace, the spectral features of amplitude and phase components, along with the down-sampled features of the upper layer, can then be properly fused together on a common ground in from the finest to the coarsest scale to derive the final latent code. The detailed design of SAFM can be found in~\cite{su2021seeing}.

The Conv-block($L$) contains two stacks of $L \times L$ convolution, batch normalization, and ReLU operations. Because the properties of the spectral bands of amplitude and phase can be significantly different, we mostly use $L=1$ to learn the best linear combination of multi-spectral features to avoid noise confusion and reduce the number of model parameters. The up-sampler and down-sampler perform $2\times$ and $\frac{1}{2} \times$ scaling, respectively. The skip connections~(SC) directly pass the feature maps of different spatial scales from the individual encoder branches
to the Chanel Fusion blocks of their corresponding branches of the decoder. 

In the decoder path,  the Channel Fusion block in each scale-branch involves a up-sampler, a Channel Attention Module (CAM), and a Conv-block. CAM aims to fuse the multi-scale features from different spectral bands in the channel dimension using the channel attention mechanism detailed in~\cite{su2021seeing}. The Conv-block has the same functional blocks as that in the encoder. Each decoding-branch receives a ``shallower-layer''  feature map from the corresponding encoding-branch via the skip-connection shortcut and concatenates the feature map with the upsampled version of the decoded ``deeper-layer''  feature map from its coarser-scale branch.  Besides, the concatenated feature map is then processed by CAM to capture the cross-channel attention to complement the local region for image restoration.

\vspace{-0.2in}
\subsection{Performance Impacts of Physical Guides on THz CT Imaging}

Herein we demonstrate the performance gain brought by the physics guides on THz computational imaging, particularly on THz image restoration and 3D CT. In the experiments, totally seven sample objects are printed, measured, and aligned for evaluation \footnote{The THz-TDS Image Dataset can be found at \href{https://github.com/wtnthu/THz\_Data}{https://github.com/wtnthu/THz\_Data}}.

To the best of our knowledge, there is no method specially designed for restoring THz images besides \texttt{Time-max}~\cite{hung2019terahertz}. Thus, we compare our method against three representative CNN-based image restoration models, including DnCNN~\cite{zhang2017beyond},  NBNet~\cite{cheng2021nbnet}, and the baseline U-Net~\cite{ronneberger2015u}, trained on the image pairs of \texttt{Time-max} images and their corresponding ground-truths.  
For objective quality assessment, we adopt two widely-used metrics including the Peak Signal-to-Noise Ratio (PSNR) and Structural SIMilarity (SSIM) to respectively measure the pixel-level and structure-level similarities between a restored image and its ground-truth as shown in Table\,\ref{tab:psnr_ssim}. We also adopt the Mean-Squared Error (MSE) between the cross-sections of a reconstructed 3D tomography and the corresponding ground-truths for assessing the 3D reconstruction accuracy as compared in Table\,\ref{tab:mse}. 

\begin{table*}[t]
		\caption{\small Quantitative comparison (PSNR and SSIM) of THz image restoration performances with different methods on seven test objects.  ($\uparrow$: higher is better)}
		\begin{center}
		    \vspace{-0.25in}
			\begin{scriptsize}
			\scriptsize
			\scalebox{0.9}{
				\begin{tabular}{|c|c|c|c|c|c|c|c|c|c|c|c|c|c|c|} \hline
				     Method & \multicolumn{7}{c|}{PSNR$\uparrow$} & \multicolumn{7}{c|}{SSIM$\uparrow$} \\
				    \cline{2-8} \cline{9-15}
					                                     &Deer   &DNA    &Box    &Eevee  &Bear    &Robot   &Skull    &Deer   &DNA   &Box  &Eevee   &Bear    &Robot   &Skull\\ \hline\hline
				    $\texttt{Time-max}$                             &12.42  &12.07  &11.97   &11.20    &11.21        &11.37   &10.69  &0.05   &0.05   &0.14  &0.14     &0.12        &0.08    &0.09 \\ \hline
					$\texttt{DnCNN-S}$ \cite{zhang2017beyond}       &19.94  &23.95  &19.13   &19.69    &19.44        &19.72   &17.33  &0.73   &0.77   &0.73  &0.72     &0.63        &0.77    &0.36 \\ \hline
					$\texttt{NBNet}$ \cite{cheng2021nbnet}          &20.24  &25.10  &20.21   &19.84    &20.12        &20.01      &19.69   &0.81   &0.85   &0.75  &0.77     &0.80        &0.80      &0.78   \\ \hline
					$\texttt{U-Net}$ \cite{ronneberger2015u}    &19.84	 &24.15  &19.77	  &19.95    &19.09        &18.80   &17.49  &0.55   &0.78   &0.77  &0.76     &0.56        &0.76    &0.51 \\ \hline
					$\texttt{SARNet}$                      &\textbf{22.98}	 &\textbf{26.05}  &\textbf{22.67}	 &\textbf{20.87}  &\textbf{21.42}   &\textbf{22.66}   &\textbf{22.48}  &\textbf{0.84}  &\textbf{0.90} &\textbf{0.83}   &\textbf{0.82}   &\textbf{0.82}   &\textbf{0.83}   &\textbf{0.84} \\ \hline
				\end{tabular}}
			\end{scriptsize}
		\end{center}
		\label{tab:psnr_ssim}
 		\vspace{-0.4in}
	\end{table*}

\begin{table*}[t]
		\caption{\small Quantitative comparison of MSE between the cross-sections of a reconstructed 3D tomography and their ground-truths with different methods on seven test objects.  ($\downarrow$: lower is better)}
		\begin{center}
		    \vspace{-0.24in}
			\begin{scriptsize}
			\scriptsize
			\scalebox{0.9}{
				\begin{tabular}{|c|c|c|c|c|c|c|c|c|c|c|c|c|c|c|} \hline
					 Method  & \multicolumn{7}{c|}{MSE$\downarrow$} \\
                    \cline{2-8}
                    					                                     &Deer   &DNA    &Box     &Eevee    &Bear    &Robot   &Skull  \\ \hline
				    $\texttt{Time-max}$                                      &0.301  &0.026  &0.178   &0.169    &0.084        &0.203   &0.225 \\ \hline
					$\texttt{DnCNN-S}$ \cite{zhang2017beyond}                &0.153  &0.162  &0.309   &0.149    &0.056        &0.223   &0.293  \\ \hline
					$\texttt{NBNet}$ \cite{cheng2021nbnet}                   &0.240  &0.184  &0.305   &0.134    &0.088        &0.128   &0.138     \\ \hline
					$\texttt{U-Net}$ \cite{ronneberger2015u}                 &0.227	 &0.166  &0.266	  &0.157    &0.077        &0.093   &0.319  \\ \hline
					$\texttt{SARNet}$         &\textbf{0.107}	 &\textbf{0.015}  &\textbf{0.041}  &\textbf{0.105}   &\textbf{0.050}   &\textbf{0.065}    &\textbf{0.052} \\ \hline
				\end{tabular}}
			\end{scriptsize}
		\end{center}
		\label{tab:mse}
 		\vspace{-0.45in}
	\end{table*}

\begin{figure}[t]
\centering
\vspace{-0.2in}
\includegraphics[width=0.9\textwidth]{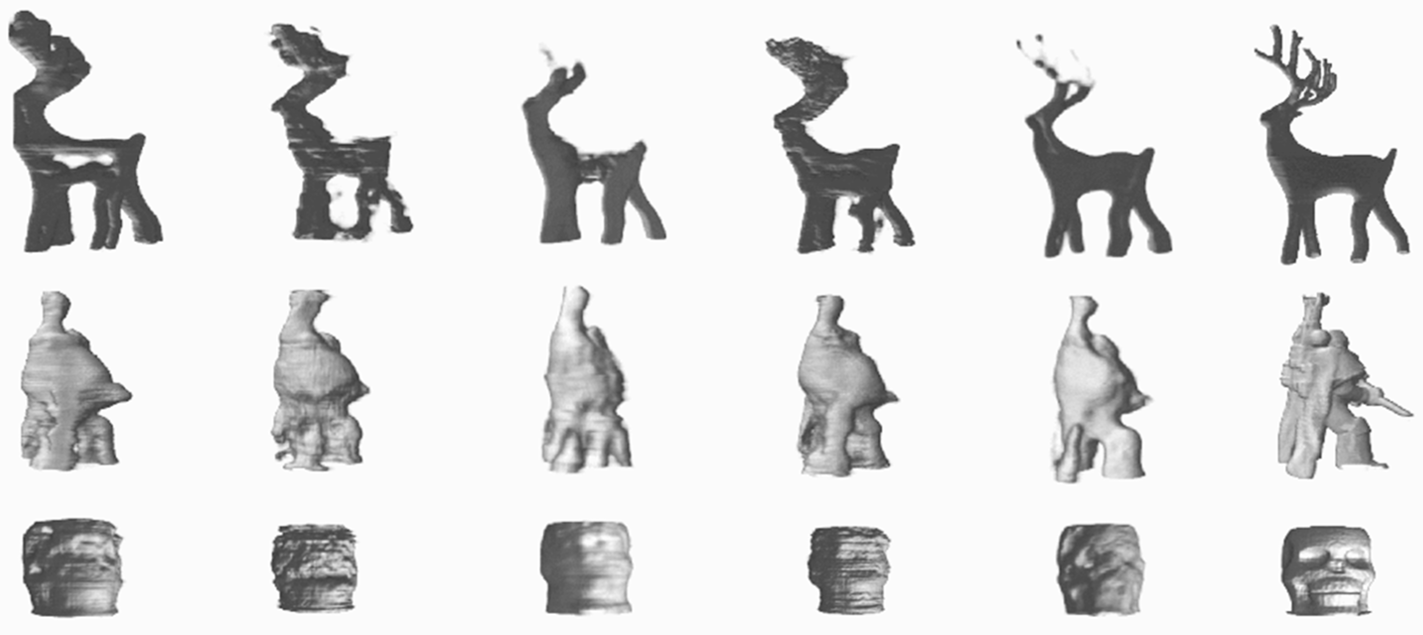}
\vspace{-0.2in}
\caption*{(a)  \ \ \ \ \ \ \ \ \ \ \ \ \ \ (b)  \ \ \ \ \ \ \ \ \ \ \ \ \ \ (c)  \ \ \ \ \ \ \ \ \ \ \ \ \ \ (d)  \ \ \ \ \ \ \ \ \ \ \ \ \ \ \ (e)  \ \ \ \ \ \ \ \ \ \ \ \ \ \ (f)}
\vspace{-0.2in}
\caption{\small Illustration of 3D CT reconstructions on \textbf{Deer}, \textbf{Robot}, and \textbf{Skull} from left to right: (a) $\texttt{Time-max}$, (b) $\texttt{DnCNN-S}$ \cite{zhang2017beyond},  (c) $\texttt{NBNet}$ \cite{cheng2021nbnet}, (d) $\texttt{U-Net}$ \cite{ronneberger2015u},  (e) $\texttt{SARNet}$,  and (f) the ground-truth.} 
\label{fig:3d}	
\vspace{-0.3in}
\end{figure}

Table\,\ref{tab:psnr_ssim} shows that the four CNN-based data-driven restoration methods  all achieve large performance gain over $\texttt{Time-max}$. Moreover, thanks to the physical guides, \texttt{SARNet} significantly outperforms the competing CNN models on all sample objects in both metrics. 
Similarly, in terms of  3D reconstruction accuracy, Table\,\ref{tab:mse} demonstrates that \texttt{SARNet} stably achieves significantly lower average MSE of tomographic reconstruction than the competing methods on all sample objects. For qualitative evaluation, Figure~\ref{fig:3d} illustrates the 3D reconstructions of \textbf{Deer}, \textbf{Robot}, and \textbf{Skull}\footnote{Complete 3D CT results on all objects can be found at \href{https://github.com/wtnthu/THz_Tomography_2022}{https://github.com/wtnthu/THz\_Tomography\_2022}}, showing that  \texttt{SARNet} reconstructs significantly clearer and  faithful 3D images with finer details such as more correct thickness of body and clear antlers of \textbf{Deer}, accurate facial contour of \textbf{Skull}, and the gun in \textbf{Robot}’s hand, achieving by far the best 3D THz tomography reconstruction quality in the literature.  Both the quantitative and qualitative evaluations confirm a significant performance leap with the physics-guided \texttt{SARNet} over the competing methods, demonstrating the high efficacy of curated physical guides on THz CT computational imaging.  

\section{Concluding Remarks}
\label{sec:remark}
THz computational imaging is an emerging research field with great application potential. This article briefly introduced the associations of physical characteristics with major THz computational imaging configurations, including compressive sensing imaging, pulse imaging, holographic imaging, and computed tomography. With THz computational imaging, multiple physical characteristics can be modeled and contribute to various applications, such as material exploration, industrial inspection, security screening, chemical inspection, and non-destructive evaluation. With the rich information in multi-dimensional THz signal behaviors, developing the different THz imaging modalities guided by physical characteristics is the key to the door of THz computational imaging. We have illustrated a THz tomographic imaging example showing how to leverage the prominent spectral information carried in THz time-domain signals, guided by the water absorption profile of THz light-matter interaction, to devise a data-driven CNN model for effectively restoring corrupted THz images. With such well-designed image processing, our experiments confirm a performance leap from the relevant state-of-the-arts in THz tomographic imaging. This also sheds light on the strength and opportunities of signal processing in THz computational imaging.
\ifCLASSOPTIONcaptionsoff
  \newpage
\fi



%
\bibliographystyle{IEEEtran}
\bibliography{ref}

%




\end{document}